\documentclass[namedreferences,hyperref,optionalrh,solaromanenum]{spr-sola}

\usepackage{graphicx}                    

\usepackage{amsmath, amssymb, bm}
\usepackage{soul}
\usepackage{longtable}
\usepackage{makecell}
\usepackage{booktabs} 
\usepackage{multirow}
\usepackage{array}
\usepackage{multicol}
\usepackage{url}
\usepackage{xurl}
\usepackage{comment}
\usepackage{lineno}
\usepackage[inline]{trackchanges}
\usepackage{natbib}
\usepackage{float}
\usepackage{lmodern}
\usepackage{graphicx}
\usepackage{listings}
\usepackage[dvipsnames]{xcolor}
\usepackage[ruled,vlined,linesnumbered,algo2e]{algorithm2e}
\usepackage{microtype}

\usepackage{dirtytalk}
\usepackage{cleveref}

\begin{document}

\begin{frontmatter}

\title{Prediction of Magnetic Flux Evolution During Solar Active Region Emergence using Long Short-Term Memory Networks}

%

\author[addressref={1},corref,email={ebd@njit.edu}]{\inits{E.}\fnm{Eren}~\snm{Dogan}\orcid{0009-0002-9903-9134}}

\author[addressref={2}]{\inits{S.}\fnm{Spiridon}~\snm{Kasapis}\orcid{0000-0002-0972-8642}}

\author[addressref={3}]{\inits{S.}\fnm{Sarang}~\snm{Patil}\orcid{0009-0005-2657-1156}}

\author[addressref={1}]{\inits{J.}\fnm{Jonas}~\snm{Tirona}\orcid{0009-0007-3333-6685}}

\author[addressref={5}]{\inits{J.}\fnm{John}~\snm{Stefan}\orcid{0000-0002-5519-8291}}

\author[addressref={4}]{\inits{I.}\fnm{Irina}~\snm{Kitiashvili}\orcid{0000-0003-4144-2270}}

\author[addressref={3}]{\inits{M.}\fnm{Mengjia}~\snm{Xu}\orcid{0000-0002-3627-8261}}

\author[addressref={4,5}]{\inits{A.}\fnm{Alexander}~\snm{Kosovichev}\orcid{0000-0003-0364-4883}}

%

\address[id={1}]{Department of Computer Science, New Jersey Institute of Technology, Newark, NJ, USA}
\address[id={2}]{Department of Astrophysical Sciences, Princeton University, Princeton, NJ, USA}
\address[id={3}]{Department of Data Science, New Jersey Institute of Technology, Newark, NJ, USA}
\address[id={4}]{NASA Advanced Supercomputing Division, NASA Ames Research Center, Moffett Field, CA, USA}
\address[id={5}]{Department of Physics, New Jersey Institute of Technology, Newark, NJ, USA}

\begin{abstract}
Solar active regions (ARs) are the primary drivers of space weather events, making their early prediction crucial for operational forecasting systems. We develop machine learning models capable of predicting the evolution of magnetic flux during AR emergence using 1D time series of the continuum intensity and solar oscillation power maps for 53 active regions and their surrounding quiet-Sun areas. Each observable is sampled over a fixed $30.66^{\circ} \times 30.66^{\circ}$ field of view. These observations capture the temporal evolution of each active region and serve as inputs for training and validation of our \texttt{MagFluxLSTM} and \texttt{MagFluxEnc-Dec} models. The \texttt{MagFluxLSTM} architecture implements a single-stage standard Long-Short Term Memory (LSTM) network. \texttt{MagFluxEnc-Dec} represents an LSTM encoder-decoder with teacher forcing. To test and evaluate the models' performance, we use the continuum intensity and oscillation power maps (calculated for several frequency bands from Doppler velocity) as input to predict the magnetic flux. Among the top 100 hyperparameter configurations ranked by validation derivative RMSE, 98\% correspond to \texttt{MagFluxLSTM}, compared to only 2\% for \texttt{MagFluxEnc-Dec}. Thus, although the \texttt{MagFluxEnc-Dec} architecture has higher model complexity, it leads to poorer generalization to ARs outside the training set and less stable training than the simpler \texttt{MagFluxLSTM}, which can predict magnetic flux emergence 3–10 hours in advance within a 12-hour prediction window in both experimental and operational-type settings for the 5 testing active regions. 
\end{abstract}



\end{frontmatter}

%

\section{Introduction} \label{sec:Introduction}

Solar active regions (ARs) are areas of intense and complex magnetic field concentration on the solar photosphere and are the primary source of space weather phenomena, including solar flares, coronal mass ejections, and solar energetic particle events \citep[e.g.,][]{schrijver2009driving}. Therefore, early detection and prediction of AR emergence are critical for space weather forecasting, as they can potentially enhance warning capabilities for geomagnetic disturbances that affect Earth's magnetosphere and impact technological infrastructure. Understanding subsurface dynamics and changes in turbulent motions that cause the stochastic excitation of solar oscillations at the photosphere, the power spectrum of which can be sensitive to subsurface perturbations caused by emerging AR \citep{hartlep2011signatures,singh2016high,waidele2023strengthening,kasapis2023predicting}. Such pre-emergence signatures can also be detected through H$\alpha$ imaging \citep{liggett1985}, surface flow analysis \citep{grigor2011dynamics,schunker2024flux}, and helioseismic travel-time variations \citep{kosovichev1999,kosovichev2011,ilonidis2011detection,stefan2023exploring}. A number of helioseismic studies have confirmed that variations of photospheric disturbances could signal the upcoming emergence of magnetic flux before it can be identified visually \citep[e.g.,][]{ilonidis2011detection, ilonidis2013helioseismic,birch2012helioseismology, leka2013helioseismology, barnes2014helioseismology,stefan2023exploring}.

Machine learning (ML) models for forecasting space weather events have recently emerged as a powerful approach for analyzing the growing volume of heliophysics data \citep{camporeale2020mlhelio}. These methods have been successfully applied to a wide range of predictive tasks, including the prediction of solar flare onset \citep[e.g,][]{jiao2020solar,georgoulis2021flare}, solar energetic particles \citep[SEPs; e.g.,][]{kasapis2026review, hutchins2026solar}, and coronal mass ejections \citep[CMEs; e.g,][]{bobra2016predicting,hegde2025predicting}. One of the most recent advances in this domain is the SuryaFM model \citep{roy2025surya,roy2025suryabench}, a 366-million-parameter foundation model for heliophysics designed to learn general-purpose solar representations from multi-instrument SDO observations.
In this light, deep learning architectures designed for sequential data offer promising avenues for capturing the temporal dependencies inherent in the emergence of active regions. Long Short-Term Memory (LSTM) networks--an advanced architecture of recurrent neural networks (RNNs)--have demonstrated good performance in time-series prediction tasks across diverse domains \citep{hochreiter1997long, hua2019deep, sherstinsky2020fundamentals}, making them particularly suitable for modeling the complex temporal evolution of solar magnetic fields. 

\citet{kasapis2023predicting, kasapis2025prediction} employed LSTM architectures to predict the decrease in continuum intensity associated with the emergence of ARs before they were observed on the solar surface. The LSTM models were trained on Solar Dynamics Observatory/Helioseismic and Magnetic Imager \citep[SDO/HMI;][]{pesnell2012solar, scherrer2012helioseismic} observations processed into the 1D ML-ready dataset, the Solar Active Region Emergence Dataset \citep[SolARED;][]{kasapis2026solared}. The LSTM models were designed to predict the onset of continuum intensity decrease associated with AR emergence within a 12-hour window. The results demonstrate that the LSTM-based analysis of oscillation power maps can successfully forecast AR emergence 5--29 hours before the associated continuum intensity decrease, at a stage when the magnetic flux has reached only 4\%--9.6\% of its eventual maximum.

This study aims to test whether ML models can capture disturbances in solar acoustic power and continuum intensity to predict the magnetic flux emergence before it is visible on the solar photosphere. The presented deep learning framework for predicting the evolution of magnetic flux builds on the methodology of \cite{kasapis2025prediction} and addresses three key challenges: a) capturing long-term temporal dependencies and pre-emergence signatures in acoustic power and continuum intensity observations, b) demonstrating the ability to predict magnetic flux given (a), and c) achieving operational magnetic flux forecast capabilities with sufficient lead time for space weather applications. The paper is organized as follows: Section~\ref{sec:method} presents the methodology followed, such as the data preprocessing steps, the LSTM architectures used, the loss optimization, and the LSTM training procedures. Section~\ref{sec:results} describes the prediction results on five testing ARs, including quantitative performance metrics and qualitative analysis of prediction capabilities. Finally, Section~\ref{sec:Discussion} provides an overview of the key findings, discusses the models' limitations, and offers some concluding remarks.

\section{Data and Methodology} \label{sec:method}

We provide detailed information about the AR emergence dataset (Section~\ref{sec:Dataset}) the different LSTM architectures (Section~\ref{sec:architecture}), and the model training (Section~\ref{sec:training_and_loss}) and implementation details (Section~\ref{sec:implementation}) in the following subsections.

\subsection{Dataset} \label{sec:Dataset}

In this study, we train and test the developed ML models on the compressed ML-ready data sets derived from preprocessed solar surface observations of 53 ARs, observed from March 2010 to June 2023 using the SDO/HMI \citep{scherrer2012hmi, hoeksema2014hmi} instrument. The dataset's 1D time-series represent the evolution of spatially averaged in $9\times 9$ bins: a) continuum intensity $I_c$, b) unsigned magnetic flux $\Phi_m$, and c) $M = 4$ frequency ranges of acoustic power $P_a$ (i.e., 2--3 mHz, 3--4 mHz, 4--5 mHz, and 5--6 mHz), all at 1-hour cadence, covering the full emergence and evolution lifecycle of each AR. The acoustic power maps are derived from an 8-hour series of Dopplergrams with one-hour shifts, capturing the temporal evolution of subsurface dynamics throughout each AR's lifecycle. A detailed description of the Solar Active Region Emergence Dataset (SolARED) is presented in \cite{kasapis2026solared}. The dataset is available through the SolARED interactive website at \url{https://sun.njit.edu/sarportal/}. The SolARED dataset utilized in this study is already pre-normalized to account for the solar geometric effect.

In the original dataset, the evolution of these observables was tracked and remapped onto the heliographic coordinates using Postel's projection centered on the center of the $30.66^{\circ} \times 30.66^{\circ}$ regions to correct for geometrical distortions. The tracked and remapped series of Dopplergrams, sampled at a 45~s cadence, are used to calculate oscillation power maps, which are then divided into 81 tiles arranged in a $9 \times 9$ grid. To focus on the primary AR emergence area and balance the active and quiet Sun tiles, the central row of the $9 \times 9$ grid is isolated, and the top and bottom four rows are excluded from this analysis. This results in a spatial selection of only 9 horizontal tiles as seen in Figure~\ref{fig:pipeline1}, similar to the tile selection done by \cite{kasapis2025prediction} and by \cite{tirona2026forecasting}. The continuum intensity $I_c$ and acoustic power $P_a$ timelines are stacked together to produce the final input tensor for each AR with dimensions $T \times N \times (M+1)$, where $T = 9$ represents the central row of tiles, $N = 240$ corresponds to the temporal extent (240 hours), and $M + 1 = 5$ denotes the acoustic power frequency bands plus one additional channel for continuum intensity. Each of the 9 tiles is treated as an independent spatial sample within the batch dimension during training. An overview of this data accumulation process can be seen in Figures~\ref{fig:pipeline1} and~\ref{fig:pipeline2}.

The training data comprises of 41 ARs (77\%), the validation data includes 7 ARs (13\%), and the testing group utilizes the same 5 ARs (10\%) evaluated in \cite{kasapis2025prediction} and \cite{tirona2026forecasting}. This split ensures that the model is evaluated on ARs outside the training set, providing a realistic assessment of generalization capability. During training, $W = 110$ hours of observations are used as inputs to the model, with the next $P = 12$ hours of observations reserved for use as targets. This 122-hour (W+P hours) window slides through the 240-hour temporal extent 118 times, generating distinct input–target pairs that are used to train the ML models in a supervised manner. For testing, the input window lengths were adjusted for each test AR to make sure that the very first input intervals capture only pre-emergence phases. While all ARs are tested using 96-hour input timelines, AR11726 is limited to 72 hours due to its emergence closer to the Eastern limb. The SolARED dataset utilized in this study is normalized to account for the solar geometric effect, with all results denormalized for presentation purposes. In addition, min-max scaling is applied independently to each input channel to avoid imposing artificial relationships between physically distinct quantities, while preventing data leakage by fitting the scaler exclusively on the training set.

\subsection{Model Architecture} \label{sec:architecture}

We investigated two LSTM-based model architectures in this study: a) an encoder–decoder LSTM (\texttt{MagFluxEnc-Dec}; Figure~\ref{fig:pipeline1}) that employs autoregressive decoding with teacher forcing and b) a stacked LSTM (\texttt{MagFluxLSTM}; Figure~\ref{fig:pipeline2}) that performs one-shot, multi-step predictions. Both models take temporal sequences of magnetic and acoustic features as input and generate $P$-timesteps–ahead forecasts of magnetic flux evolution. As mentioned above, the output length was selected to be $P = 12$ hours, but predictions with shorter or longer output windows are also possible. 

\subsubsection{\texttt{MagFluxEnc-Dec}: Encoder-Decoder LSTM}
\begin{figure}[htbp]
\centerline{\includegraphics[width=\textwidth,clip=]{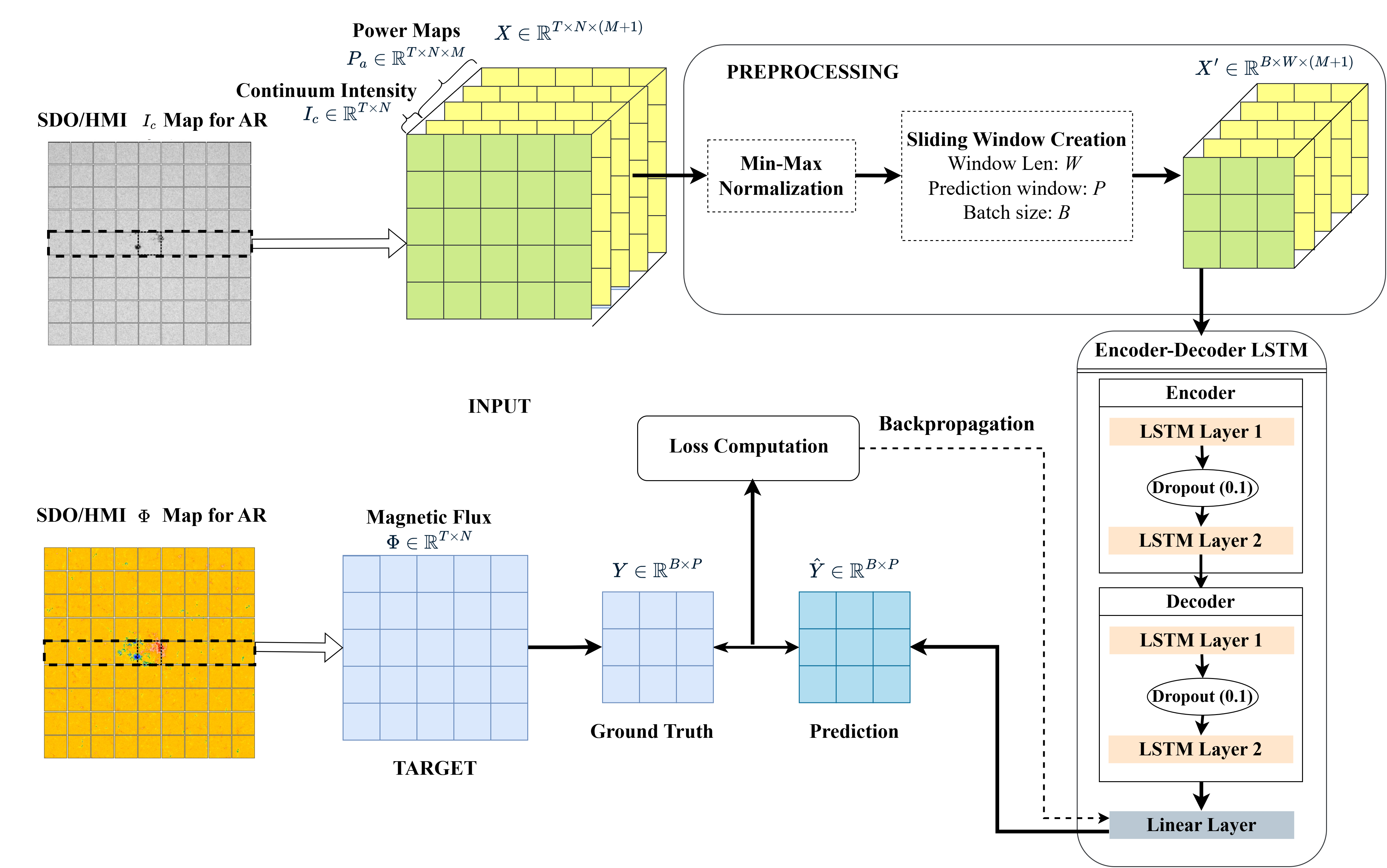}}
\caption{The \texttt{MagFluxEnc-Dec} pipeline for magnetic flux prediction. The system extracts central tiles from SDO/HMI continuum intensity ($I_c$) and magnetic flux ($\Phi_m$) maps, combines them with acoustic power maps to form feature tensor $X \in {R}^{T \times N \times (M+1)}$ where $T=9$ central tiles, $N=240$ timesteps, and $M=4$ power map frequencies. Sliding windows are generated by continuously sliding a window of length $W = 110$ timesteps across the temporal dimension with unit stride, extracting the following $P = 12$ timesteps as prediction targets, yielding a batch tensor $X' \in {R}^{B \times W \times (M+1)}$. A two-layer encoder-decoder LSTM is employed to predict magnetic flux evolution $\hat{Y} \in {R}^{B \times P}$.}
\label{fig:pipeline1}
\end{figure}

The \texttt{MagFluxEnc-Dec} model implements an encoder-decoder LSTM architecture, which is considered the state-of-the-art approach for sequence-to-sequence prediction tasks with variable-length temporal dependencies. The architecture consists of two symmetric components: a multi-layer LSTM encoder that compresses the input window into a fixed-dimensional context state, and a recurrent LSTM decoder that autoregressively generates the future sequence using this context (see Figure~\ref{fig:pipeline1}).

The encoder processes the input sequence $\mathbf{x}_{1:W}$ of length $W = 110$ timesteps via a multi-layer LSTM with $L$ stacked layers and $h$ hidden units per layer. The number of layers $L$ is treated as a dynamic hyperparameter optimized during the training search process. At each encoder timestep $t = 1, \ldots, W$, the recurrence relations are:
\begin{equation}
\begin{aligned}
\mathbf{H}_t^{(1)}, \mathbf{c}_t^{(1)} &= \mathrm{LSTM}^{(1)}_{\text{enc}}\left(\mathbf{x}_t,\ \mathbf{H}_{t-1}^{(1)},\ \mathbf{c}_{t-1}^{(1)}\right), \\
\vdots \\
\mathbf{H}_t^{(l)}, \mathbf{c}_t^{(l)} &= \mathrm{LSTM}^{(l)}_{\text{enc}}\left(\mathbf{H}_t^{(l-1)},\ \mathbf{H}_{t-1}^{(l)},\ \mathbf{c}_{t-1}^{(l)}\right),
\end{aligned}
\label{eq:encoder-seq2seq}
\end{equation}
where \(\mathbf{x}_t\) is the input feature vector, and \(\mathbf{H}_t^{(l)}\) and \(\mathbf{c}_t^{(l)}\) are the hidden state and cell state of the encoder layer \(l \in \{1, \ldots, L\}\) at timestep \(t\). The final encoder hidden and cell states summarize the entire window history and are transferred to initialize the decoder. The decoder adopts the same multi-layer LSTM architecture as the encoder, and autoregressively predicts the output sequence $\hat{y}_{1:P}$ over $P = 12$ future timesteps. For each prediction timestep $j = 1, \ldots, P$, the decoder transitions according to:
\begin{equation}
\begin{aligned}
\mathbf{H}_j^{(1)}, \mathbf{c}_j^{(1)} &= \mathrm{LSTM}^{(1)}_{\text{dec}}\left(z_j,\ \mathbf{H}_{j-1}^{(1)},\ \mathbf{c}_{j-1}^{(1)}\right), \\
\vdots\\
\mathbf{H}_j^{(l)}, \mathbf{c}_j^{(l)} &= \mathrm{LSTM}^{(l)}_{\text{dec}}\left(\mathbf{H}_j^{(l-1)},\ \mathbf{H}_{j-1}^{(l)},\ \mathbf{c}_{j-1}^{(l)}\right), \quad l = 2, \ldots, L, \\
\hat{y}_j &= \mathbf{W}_{\text{out}} \mathbf{H}_j^{(L)} + b_{\text{out}},
\end{aligned}
\label{eq:decoder-seq2seq}
\end{equation}
where \(z_j\) is the input to the decoder at prediction timestep \(j\), determined by the teacher forcing mechanism. \(\mathbf{H}_j^{(l)}\) and \(\mathbf{c}_j^{(l)}\) are the hidden and cell states of decoder layer \(l\) at timestep \(j\). The hidden state of the final decoder layer $L$, \(\mathbf{H}_j^{(L)}\), is projected via the weight matrix \(\mathbf{W}_{\text{out}}\) and bias \(b_{\text{out}}\) to produce \(\hat{y}_j\), the predicted magnetic flux value at timestep \(j\). Specifically, $z_j$ is determined by the \emph{teacher forcing} mechanism: during training, with probability $P_{tf}$ (the teacher forcing ratio), $z_j$ is set to the ground-truth target value $y_{j-1}$; otherwise, $z_j$ is set to the model's own previous prediction $\hat{y}_{j-1}$. This stochastic switching between ground-truth and predicted inputs stabilizes training by mitigating exposure bias and improving convergence. The encoder-decoder framework explicitly models how prediction errors can propagate through the iterative prediction window, making it well-suited for capturing non-stationary, autoregressive temporal dynamics such as those present in solar magnetic flux emergence.

\subsubsection{\texttt{MagFluxLSTM}: Stacked LSTM}

\begin{figure}[htbp]
\centerline{\includegraphics[width=\textwidth,clip=]{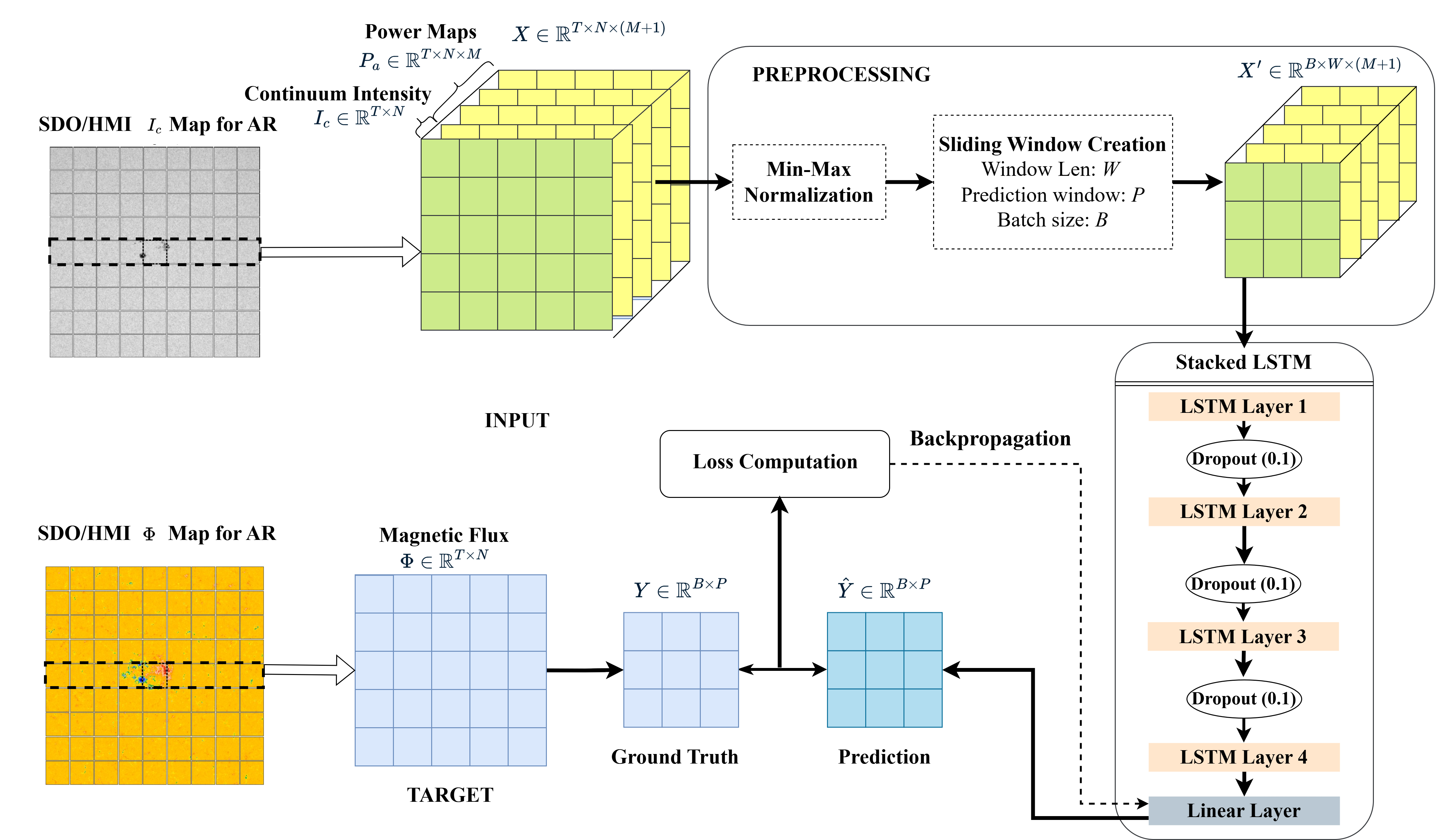}}
\caption{Schematic overview of the \texttt{MagFluxLSTM} pipeline for magnetic flux prediction during active region emergence. \texttt{MagFluxLSTM} employs stacked LSTM structure instead of the encoder-decoder structure in Figure~\ref{fig:pipeline1}. The system extracts central tiles from SDO/HMI continuum intensity ($I_c$) and magnetic flux ($\Phi_m$) maps, combines them with oscillation power maps to form feature tensor $X \in {R}^{T \times N \times (M+1)}$ where $T=9$ central tiles, $N=240$ timesteps, and $M=4$ power map frequencies. Sliding windows are generated by continuously sliding a window of length $W = 110$ timesteps across the temporal dimension with unit stride, extracting the following $P = 12$ timesteps as prediction targets. During training, windowed samples are grouped into batches of size $B$, yielding a batch tensor $X' \in {R}^{B \times W \times (M+1)}$. A four-layered stacked LSTM is employed to predict magnetic flux evolution $\hat{Y} \in {R}^{B \times P}$.}
\label{fig:pipeline2}
\end{figure}

The \texttt{MagFluxLSTM} model implements a multi-layer stacked LSTM that processes the entire input window in a single forward pass and produces all $P$-step future predictions via a single linear output layer. Unlike the encoder-decoder approach, this architecture does not employ autoregressive feedback; instead, all future timesteps are jointly decoded from the final hidden state of the stacked layers (Figure~\ref{fig:pipeline2}).

Given an input sequence $\mathbf{X} = (\mathbf{x}_1, \ldots, \mathbf{x}_W)$ of length $W = 110$, the state of the stacked LSTM, with $L$ layers (where $L$ is optimized via the Ray Tune~\citep{liaw2018tune}) and hidden size $h$, evolves as follows. At each timestep $t = 1, \ldots, W$:
\begin{equation}
\begin{aligned}
\mathbf{H}_t^{(1)}, \mathbf{c}_t^{(1)} &= \mathrm{LSTM}^{(1)}\left(\mathbf{x}_t,\ \mathbf{H}_{t-1}^{(1)},\ \mathbf{c}_{t-1}^{(1)}\right), \\
\vdots \\
\mathbf{H}_t^{(l)}, \mathbf{c}_t^{(l)} &= \mathrm{LSTM}^{(l)}\left(\mathbf{H}_t^{(l-1)},\ \mathbf{H}_{t-1}^{(l)},\ \mathbf{c}_{t-1}^{(l)}\right), \quad l = 2, \ldots, L,
\end{aligned}
\label{eq:stacked-lstm}
\end{equation}
where, $\mathbf{x}_t$ is the input feature vector at timestep $t$. 

Between consecutive LSTM layers (i.e., the output of layer $l$ and the input to layer $l+1$), dropout with rate $P_d $ is applied to improve regularization and reduce overfitting. After processing the entire window, the final hidden state of the top layer $\mathbf{H}_W^{(L)}$ encodes a comprehensive summary of the historical temporal dynamics. This summary is then projected directly to the prediction horizon via:
\begin{equation}
\hat{\mathbf{y}} = \mathbf{W}_{\text{out}} \mathbf{H}_W^{(L)} + \mathbf{b}_{\text{out}} \in \mathbb{R}^{P},
\label{eq:vanilla-output}
\end{equation}
where $\mathbf{W}_{\text{out}} \in \mathbb{R}^{P \times h}$ and $\mathbf{b}_{\text{out}} \in \mathbb{R}^{P}$ are trainable parameters, and $\hat{\mathbf{y}} = (\hat{y}_1, \ldots, \hat{y}_P)$ is the complete prediction vector for the next $P = 12$ hours. Critically, this model is \emph{non-autoregressive}: the prediction for timestep $j$ does not depend on the model's prediction at timestep $j-1$, nor is there teacher forcing. Instead, all $P$ predictions are determined simultaneously and independently by the single context vector $\mathbf{H}_W^{(L)}$. This simplicity offers computational efficiency and can outperform autoregressive models when the prediction horizon is relatively short ($P = 12$ hours), and temporal dependencies are not strongly autoregressive in nature. 

\subsection{Training and Loss Functions} \label{sec:training_and_loss}

To define magnetic flux emergence, we adopt an emergence criterion similar to that of \citet{kasapis2025prediction}, where emergence is identified when the time derivative of the $\Phi_m$ unsigned magnetic flux exceeds a fixed $\theta$ value threshold for a sustained duration of $t_{sus}$:
\begin{equation}
\begin{split}
\frac{d\Phi_m}{dt} > 0.01 \quad
\text{for} \quad t_{sus} = 4\quad {\rm hours}.
\end{split}
\label{eq:emergence_criterion}
\end{equation}

This definition is used to evaluate the timing of AR emergence in the LSTM predictions. Because the AR emergence is characterized by changes in temporal gradients rather than absolute magnitudes, a hybrid loss function is designed that explicitly incorporates derivative information to be used alongside the typical mean-squared error (MSE) loss. While magnitude loss functions are effective for predicting the true value for most purposes, the goal of this study is to predict magnetic flux growth using a time derivative. To address this goal, a hybrid loss function that simultaneously optimizes for flux magnitude accuracy and temporal derivative prediction is introduced:

\begin{equation}
\label{eq:hybrid_loss}
\begin{aligned}
\mathcal{L}_{\text{hybrid}}
&= \alpha\,\mathcal{L}_{\text{mse}} + (1-\alpha)\,\mathcal{L}_{\text{derivative}}, \quad \alpha\in(0,1].\\
\text{where} \quad
\mathcal{L}_{\text{mse}}
&= \frac{1}{BP}\sum_{b=1}^{B}\sum_{t=1}^{P}\bigl(\hat{y}_{b,t}-y_{b,t}\bigr)^2, \\
\mathcal{L}_{\text{derivative}}
&= \frac{1}{B(P-1)}\sum_{b=1}^{B}\sum_{t=2}^{P}\bigl[(\Delta \hat{y}_{b,t})-(\Delta y_{b,t})\bigr]^2, 
\end{aligned}
\end{equation}
\begin{equation}
\label{eq:loss_components}
\begin{aligned}
\\[4pt]
\Delta \hat{y}_{b,t} &= \hat{y}_{b,t}-\hat{y}_{b,t-1}\quad (t=2,\dots,P), \\
\Delta y_{b,t} &= y_{b,t}-y_{b,t-1}\quad (t=2,\dots,P), \\[4pt]
\end{aligned}
\end{equation}

The training objective in Equation~\ref{eq:hybrid_loss} combines mean-squared error (MSE) terms defined in Equation~\ref{eq:loss_components}. The first term, $\mathcal{L}_{\text{mse}}$, penalizes absolute discrepancies between predicted magnetic flux $\hat{y}$ and the target $y$ across the prediction horizon. The second term, $\mathcal{L}_{\text{derivative}}$, applies the same penalty to the first-order finite differences ($\Delta \hat{y}$ and $\Delta y$), thereby encouraging agreement in temporal gradients and improving sensitivity to the onset and timing of flux emergence. The normalization factors ($1/BP$ and $1/B(P-1)$) ensure that both components are averaged over their respective time indices, matching the common \texttt{reduction=`mean'} behavior used in practice. The mixing coefficient $\alpha\in(0,1]$ controls the trade off between amplitude fidelity ($\mathcal{L}_{\text{value}}$) and temporal-dynamics fidelity ($\mathcal{L}_{\text{derivative}}$), with smaller $\alpha$ emphasizing timing accuracy and larger $\alpha$ emphasizing magnitude accuracy. 

Because the input data are sampled at a fixed cadence of $\Delta t = 1$ hour, the finite difference terms directly approximate temporal derivatives; therefore, any constant rescaling of $\Delta t$ can be absorbed into $\alpha$ without affecting optimization. In practice, an overall constant multiplier may also be applied to $\mathcal{L}_{\text{hybrid}}$ to stabilize gradient magnitudes during training. The training workflow for the LSTM models follows a standard temporal learning pipeline with sliding-window sampling and backpropagation through time as summarized in Algorithm \ref{alg:lstm_training}. After completing all training forward passes, the model is evaluated on the validation set during inference, and the validation loss --- computed as the RMSE of the derivative --- is used to update the learning-rate scheduler. This process is repeated across all epochs.

\begin{algorithm2e}
\caption{Training and Validation Procedures}
\label{alg:lstm_training}
\SetAlgoLined
\DontPrintSemicolon

\KwIn{Training data, validation data, hyperparameters}
\KwOut{Trained LSTM model parameters}

Initialize LSTM, Adam optimizer, learning rate scheduler\;

Preprocess data: min-max normalize, create sliding windows\;

Create batches: train (shuffle/no shuffle), validation (no shuffle)\;

\For{epoch $= 1$ \KwTo $N_{\text{epochs}}$}{
    total\_train\_loss $\leftarrow 0$\;

    \ForEach{(input\_seq, target\_seq) in train\_batches}{
        Forward pass in train mode\;
        Compute loss\;
        Clip gradients, update parameters\;
        Accumulate train loss\;
    }

    Train loss $\leftarrow$ average(total\_train\_loss)\;

    total\_val\_loss $\leftarrow 0$\;

    \ForEach{(input, target) in val\_batches}{
        Forward pass in eval mode\;
        Compute derivative RMSE loss\;
        Accumulate validation loss\;
    }

    Validation loss $\leftarrow$ RMSE of prediction differences\;
    Update scheduler with validation loss\;
}
\end{algorithm2e}


\subsection{Implementation Details} 
\label{sec:implementation}

The complete ML framework discussed is implemented in PyTorch 2.7.0, leveraging CUDA acceleration for efficient training on NVIDIA GPUs. Model optimization employs the Adam optimizer with weight decay ($D_w$), and an initial learning rate of ($lr$) selected through systematic hyperparameter optimization using Ray Tune~\citep{liaw1807tune} with the Asynchronous Successive Halving Algorithm \citep[ASHA;][]{li2020system} scheduler combined with Hyperopt search~\citep{bergstra2013hyperopt}. Learning rate scheduling is implemented using \textit{ReduceLROnPlateau}\footnote{\url{https://docs.pytorch.org/docs/stable/generated/torch.optim.lr_scheduler.ReduceLROnPlateau.html}} with a decay factor of 0.2, and a grace period of 10 epochs without validation improvement, allowing adaptive learning rate reduction when training plateaus and preventing overfitting while ensuring the model converges to optimal emergence detection capability. 

Training is conducted for a maximum of 100 epochs with early stopping based on validation temporal-gradient RMSE, by the \textit{ASHAScheduler}\footnote{\url{https://docs.ray.io/en/latest/tune/api/doc/ray.tune.schedulers.ASHAScheduler.html}} and custom \textit{TrialPlateauStopper}\footnote{\url{https://docs.ray.io/en/latest/tune/api/doc/ray.tune.stopper.TrialPlateauStopper.html}} (more details are provided in \ref{appendix:hyperopt}). Training and validation datasets are batched and shuffled as determined by the \textit{Hyperopt}\footnote{\url{https://hyperopt.github.io/hyperopt/}} search. The baseline training procedure was kept consistent across all experiments. Batching was used to reduce GPU memory demands by training on smaller data subsets, while shuffling was applied to prevent memorization of AR-specific patterns and to promote learning of generalizable temporal dynamics across different active regions. Remember that in order to balance active tiles with quiet tiles and reduce noise overhead, only the middle row of the $9 \times 9$ AR grid was used as input, and the top and bottom four rows of tiles were left out.

\begin{table}[ht]
\centering
\caption{Model parameters and corresponding best-performing configurations.
Notation: $P_d$ is dropout probability; $H$ is number of hidden layers; $L$ is number of LSTM layers;
$B$ is batch size; $lr$ is learning rate; $\lambda$ is weight decay;
$\alpha$ is hybrid loss mixing coefficient; $P_{\mathrm{tf}}$ is teacher forcing ratio.
All values correspond to the optimal configuration selected after extensive
hyperparameter optimization. The top-100 refers to the percentage of times the specific model setups showed up in the top 100 models during trials. The bolded model indicates the best performing model chosen to discuss in Section~\ref{sec:results}.}
\label{table:model_comp}

\resizebox{\textwidth}{!}{%
\begin{tabular}{lcccccccccc}
\hline
\multicolumn{3}{c}{\textbf{Model setup}} & \multicolumn{8}{c}{\textbf{Hyperparameters}}\\
\cmidrule(lr){1-3}\cmidrule(lr){4-11}
\textbf{Model} & \textbf{Loss Type}  & \textbf{top-100} & $P_d$ & $H$ & $L$ & $B$ & $lr$ & $\lambda$ & $\alpha$ & $P_{tf}$\\
\midrule
\textbf{\texttt{MagFluxLSTM}}   & \textbf{hybrid loss} & \textbf{89\%} & \textbf{0.1} & \textbf{128} & \textbf{4} & \textbf{64} & $\bm{7.06\times10^{-3}}$ & $\bm{1.4\times10^{-6}}$ & \textbf{0.7} & \textbf{-} \\
\texttt{MagFluxLSTM}   & MSE loss  & 9\% & 0.2 & 64  & 1 & 32 & $2.32\times10^{-3}$ & $1.2\times10^{-5}$ & 1 & - \\
\texttt{MagFluxEnc-Dec}  & MSE loss & 0\%  & 0.3 & 32  & 4 & 32 & $1.7\!\times\!10^{-3}$ & $4.6\times10^{-6}$ & 1 & 0.5 \\
\texttt{MagFluxEnc-Dec}  & hybrid loss & 2\% & 0.2 & 64  & 4 & 32 & $9.7\!\times\!10^{-3}$ & $2.5\times10^{-5}$ & 0.9 & 0.15 \\
\hline
\end{tabular}%
}
\end{table}

The best hyperparameters for each model setup chosen by the \textit{Hyperopt} search are shown in Table~\ref{table:model_comp}. The model setup involves variations of model type and loss type. At the same time, the top-100 column highlights the percentage of times the model and loss combination was seen in the top 100 results (with different hyperparameters). The best hyperparameters are tabulated on the right-hand side of the table. The best model from this training setup was the \texttt{MagFluxLSTM} with a hybrid loss and the given hyperparameters tabulated in Table~\ref{table:model_comp}; therefore, that configuration will be discussed in more detail in Section~\ref{sec:results}.

Across our hyperparameter optimization runs, the \texttt{MagFluxLSTM} trained with the hybrid loss function emerged as the top-performing model, surpassing all other configurations. The results show that the \textit{Hyperopt} search strongly favors a) the \texttt{MagFluxLSTM} architecture over the \texttt{MagFluxEnc-Dec} model, and b) the hybrid loss over MSE loss functions. The top 100 model configurations, ranked by validation temporal-gradient RMSE of the magnetic flux, are presented in Table~\ref{table:model_comp}. Notably, the \texttt{MagFluxEnc-Dec} model trained with MSE loss does not appear among the top-performing configurations. Overall, these results indicate that the \texttt{MagFluxLSTM} with a temporal-gradient–aware loss formulation more reliably captures time-dependent behavior, whereas the \texttt{MagFluxEnc-Dec} exhibits higher variance and weaker generalization under limited training data.

\section{Results} 
\label{sec:results}

\begin{figure}[htbp]
\centerline{\includegraphics[width=\textwidth,clip=]{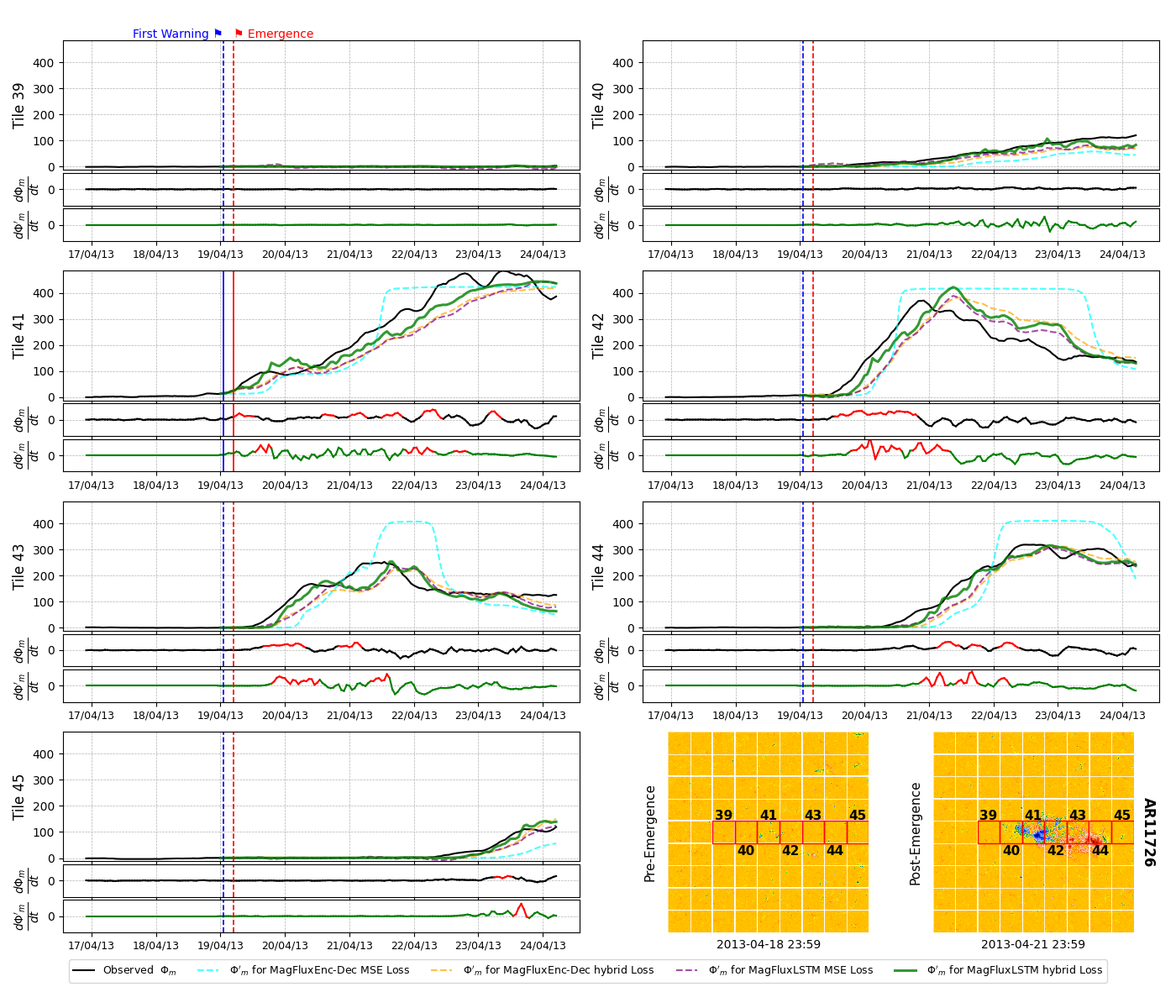}}
    \caption{The $\Phi_m$ predictions for AR11726 obtained by the models presented in Table~\ref{table:model_comp} for a 12-hour-ahead prediction window, in normalized Gauss units. The green solid line marks the \texttt{MagFluxLSTM} with the hybrid loss model, while the dashed lines show the other models. The observed flux is shown in black. The two panels below the main plot present the derivatives for a) the observed magnetic flux and b) the predicted magnetic flux by \texttt{MagFluxLSTM} hybrid loss model (top and bottom, respectively). Red segments in the derivative denote the times when the emergence criterion in Equation~\ref{eq:emergence_criterion} was fulfilled. Blue vertical dashed lines mark the first predicted emergence across all tiles, whereas the red vertical dashed lines mark the first observed emergence across all tiles. The solid vertical lines indicate the tile in which these first emergence and warning occurred, within the entire AR.} 
\label{fig:results}
\end{figure}

The results for the best-performing \texttt{MagFluxLSTM} with hybrid loss (solid line) along with the other models (dashed lines), validated using the 5 testing ARs, are shown in Figure~\ref{fig:results} (AR11726) of the main text and Figures~\ref{fig:results11698}, \ref{fig:results13165}, \ref{fig:results13179} and \ref{fig:results13183} of the \ref{appendix:results_denormalized} (ARs 11698, 13165, 13179 and 13183, respectively). Note that the model predicts magnetic flux values in the min-max scaled space, which are subsequently denormalized for visualization purposes, as seen in Figure~\ref{fig:results}. The timelines plotted in Figures~\ref{fig:results}--\ref{fig:results13183} are corrected for the solar geometric effect, and therefore the values are expressed in geometrically normalized Gauss units. Table~\ref{table:best_vanillalstm_hybrid} provides a summary of the results visualized in the aforementioned figures.

Out of the seven AR11726 tiles presented in Figure~\ref{fig:results}, the observed magnetic flux (dark solid line) in five of them (Tiles 41-45) fulfilled the criterion of Equation~\ref{eq:emergence_criterion}. The bottom right part of Figure~\ref{fig:results} shows the positions of the tiles in the original $9 \times 9$ grid, before (left; 23:59 UTC, 04-18-2013) and after (right; 23:59 UTC, 04-21-2013) AR11726 emerged. These specific tiles were chosen for demonstration since most of the activity in the tracked $512 \times 512$ pixel cutout of the solar surface is centered around them. Note that the ML model's results in Figure~\ref{fig:results} (all $\Phi_m$ timelines except the solid black) represent a value predicted 12 hours ahead. The same applies to the $d \Phi_m'/dt$ timeline, at the bottom panel of each tile's plots, which represents the derivative of $\Phi_m$ that \texttt{MagFluxLSTM} with the hybrid loss model predicted. The \texttt{MagFluxLSTM} model successfully predicts that there will be emergence in the five tiles where emergence is actually observed, while for the two quiet tiles, the model predictions never fulfilled the emergence criterion, and therefore, an emergence alarm was rightfully never produced for these tiles. The \texttt{MagFluxLSTM} Hybrid Loss model achieves an average RMSE of 27.7 across all tiles in Figure~\ref{fig:results}.

Across the testing ARs, the \texttt{MagFluxLSTM} consistently captures the rise in magnetic flux relative to each tile’s pre-emergence baseline. Table~\ref{table:best_vanillalstm_hybrid} summarizes the predicted versus observed magnetic flux increase times across the selected tiles. For AR11726 (Figure~\ref{fig:results}), the first magnetic flux increase occurs in Tile 41, marked by the red vertical line (First Emergence) as defined by the criterion in Equation~\ref{eq:emergence_criterion}. \texttt{MagFluxLSTM} successfully anticipates this emergence 4 hours in advance, which is the first predicted emergence as marked by the blue vertical line in Figure~\ref{fig:results}. This value is prescribed under the \textit{Emergence} column in Table~\ref{table:best_vanillalstm_hybrid}. In tiles 42-45, the model predicts an increase in $\Phi_m$ 6, 9, 18, and 5 hours before the observed increase, respectively (Table~\ref{table:best_vanillalstm_hybrid}, $\Phi_m$ Increase). Remember that these alarms are defined as the difference between the timestamps of the first red points in the derivative panels of the figure, plus 12 hours, to account for the fact that the $d\Phi_m'/dt$ values are provided to us by \texttt{MagFluxLSTM}, 12 hours ahead. For tiles (39 and 40), \texttt{MagFluxLSTM} with hybrid loss rightfully does not predict emergence.

Our hyperparameter sweep reveals a notable finding: the comparatively simple \texttt{MagFluxLSTM} architecture consistently surpasses the more elaborate \texttt{MagFluxEnc-Dec} model (yellow and cyan dotted lines in Figure~\ref{fig:results}). This preference can be attributed to the reduced model complexity and parameter count, leading to better generalization on the limited dataset of 53 ARs. Given a limited number of available ARs in SolARED, the lower-parameter \texttt{MagFluxLSTM} model avoids overfitting. The autoregressive decoder in \texttt{MagFluxEnc-Dec} introduces additional degrees of freedom and longer gradient pathways, potentially increasing the sensitivity to noise in the magnetic flux magnitude and derivative targets, negatively affecting the models' performance (Table~\ref{table:model_comp}). This may explain the higher ranking of the simpler model by Hyperopt: among the top 100 configurations ranked by validation derivative RMSE, 89\% correspond to \texttt{MagFluxLSTM} with hybrid loss, compared to 2\% for \texttt{MagFluxEnc-Dec}. Table~\ref{table:best_vanillalstm_hybrid} highlights the results for \texttt{MagFluxLSTM} (bold in Table~\ref{table:model_comp}). The ability of \texttt{MagFluxLSTM} with hybrid loss to predict not only the AR emergence, but also the increase in $\Phi_m$ in separate tiles, better than the rest models we experimented on, is also evident in the results of the rest validation ARs, as seen in Figures~\ref{fig:results11698}-\ref{fig:results13183} of \ref{appendix:results_denormalized}. Table~\ref{table:best_vanillalstm_hybrid} summarizes the \texttt{MagFluxLSTM} prediction results for all validation ARs.

Overall, for all validation ARs, \texttt{MagFluxLSTM} with hybrid loss produced a positive emergence alarm (the fulfillment of the Equation~\ref{eq:emergence_criterion} criterion on the predicted values), varying from 3 to 10 hours. Note that these values were obtained in an experimental setting. As described in \cite{kasapis2025prediction}, a 4-hour delay can be applied to the prediction times to account for the time required to process data in real-time. In Table~\ref{table:best_vanillalstm_hybrid}, only the $\ge 4$ hour alarms are considered operationally successful --- and are therefore marked in green, while the tiles where \texttt{MagFluxLSTM} incorrectly produced a $\Phi_m$ alarm are marked in red (ILAP; incorrect local activity predictions). When this operational criterion is applied to the magnetic flux increase predictions of the first tile emergence is observed in all validation ARs, we conclude that 3 out of five AR emergences would be predicted if the \texttt{MagFluxLSTM} with hybrid loss models were deployed in operations. This study, therefore, presents the first ML model that can correctly predict the time and space where AR magnetic flux emergence should be expected

\begin{table}[ht]
\renewcommand{\arraystretch}{1.5}
\centering
\caption{Prediction of the magnetic flux increase for the five validation ARs, for 7 select tiles. In green are the operationally successful alarms and the tiles where the \texttt{MagFluxLSTM} with hybrid loss correctly predicted that $\Phi_m$ would not increase (marked as Quiet). In orange are the alarms that would be delayed in an operational-type setting. An ILAP (incorrect local activity prediction) indicates that the model predicted an increase in $\Phi_m$ when in reality there was none.}
\label{table:best_vanillalstm_hybrid}

\resizebox{\textwidth}{!}{%
\begin{tabular}{@{}c *{7}{c} c@{}}
\hline
\textbf{AR} & \multicolumn{7}{c}{\textbf{$\Phi_m$ Increase}} & \textbf{Emergence} \\
\cmidrule(r){1-1} \cmidrule(lr){2-8} \cmidrule(l){9-9}
AR11698 & \shortstack{Tile 48\\{\color{OliveGreen}Quiet}} & \shortstack{Tile 49\\{\color{orange}4h Alarm}} & \shortstack{Tile 50\\{\color{red}ILAP}} & \shortstack{Tile 51\\{\color{OliveGreen}14h Alarm}} & \shortstack{Tile 52\\{\color{OliveGreen}13h Alarm}} & \shortstack{Tile 53\\{\color{OliveGreen}Quiet}} & \shortstack{Tile 54\\{\color{OliveGreen}Quiet}} & \shortstack{ \\{\color{OliveGreen}7h Alarm}} \\ 
\midrule
AR11726 & \shortstack{Tile 39\\{\color{OliveGreen}Quiet}} & \shortstack{Tile 40\\{\color{OliveGreen}Quiet}} & \shortstack{Tile 41\\{\color{orange} 4h Alarm}} & \shortstack{Tile 42\\{\color{OliveGreen}6h Alarm}} & \shortstack{Tile 43\\{\color{OliveGreen}9h Alarm}} & \shortstack{Tile 44\\{\color{OliveGreen}18h Alarm}} & \shortstack{Tile 45\\{\color{OliveGreen}5h Alarm}} & \shortstack{ \\{\color{orange}4h Alarm}} \\ 
\midrule
AR13165 & \shortstack{Tile 30\\{\color{OliveGreen}Quiet}} & \shortstack{Tile 31\\{\color{red}ILAP}} & \shortstack{Tile 32\\{\color{orange}4h Alarm}} & \shortstack{Tile 33\\{\color{OliveGreen}7h Alarm}} & \shortstack{Tile 34\\{\color{OliveGreen}Quiet}} & \shortstack{Tile 35\\{\color{OliveGreen}Quiet}} & \shortstack{Tile 36\\{\color{OliveGreen}Quiet}} & \shortstack{ \\{\color{OliveGreen}7h Alarm}}\\ 
\midrule
AR13179 & \shortstack{Tile 39\\{\color{OliveGreen}Quiet}} & \shortstack{Tile 40\\{\color{OliveGreen}8h Alarm}} & \shortstack{Tile 41\\{\color{orange}3h Alarm}} & \shortstack{Tile 42\\{\color{orange}1h Alarm}} & \shortstack{Tile 43\\{\color{OliveGreen}Quiet}} & \shortstack{Tile 44\\{\color{OliveGreen}Quiet}} & \shortstack{Tile 45\\{\color{OliveGreen}Quiet}} & \shortstack{ \\{\color{orange}3h Alarm}} \\ 
\midrule
AR13183 & \shortstack{Tile 39\\{\color{OliveGreen}Quiet}} & \shortstack{Tile 40\\{\color{OliveGreen}Quiet}} & \shortstack{Tile 41\\{\color{red}ILAP}} & \shortstack{Tile 42\\{\color{OliveGreen}8h Alarm}} & \shortstack{Tile 43\\{\color{red}ILAP}} & \shortstack{Tile 44\\{\color{OliveGreen}Quiet}} & \shortstack{Tile 45\\{\color{OliveGreen}Quiet}} & \shortstack{ \\{\color{OliveGreen}10h Alarm}}\\ 
\hline
\end{tabular}%
}
\end{table}

\section{Discussion and Conclusions} 
\label{sec:Discussion}

This study presents the development of a deep learning framework to predict the increase in magnetic flux associated with the emergence of ARs, addressing critical needs in space weather forecasting. Compared with earlier attempts to use an LSTM-based architecture for emergence prediction \citep{kasapis2025prediction}, this study explored several alternative approaches. First, min–max normalization is now computed strictly from the training set and applied to the validation and test sets, eliminating the information leakage present in the previous pipeline, where scaling was fit to all data. Training was also significantly transformed for traditional techniques: instead of training on each tile and active region sequentially, our model is trained in mini-batches (with optional shuffling), which yields smoother gradient updates and reduces overfitting to AR-specific patterns, such as rapid increases in magnetic flux that are not associated with emergence. Hyperparameter tuning is now performed systematically using Hyperopt and Ray Tune with ASHA early stopping, enabling exploration of a broader and more expressive model space. 

Architecturally, we extend the encoder–decoder LSTM used in prior work \citep{kasapis2025prediction} by implementing teacher forcing (\texttt{MagFluxEnc-Dec}), and introduce a simpler stacked LSTM (\texttt{MagFluxLSTM}) that ultimately outperforms \texttt{MagFluxEnc-Dec} on derivative-based validation metrics. This new validation metric uses derivative RMSE to align the optimization objective with the scientific goal of accurately predicting the time of flux emergence. Finally, we incorporate a hybrid loss function that optimizes magnitude and gradient agreement together, addressing the prior model’s inability to explicitly learn derivative dynamics. Excluding methodological differences, the biggest difference from \citet{kasapis2025prediction} is the use of continuum intensity and acoustic power variations as inputs and magnetic flux as the output. Given that a significant increase in magnetic flux precedes a decrease in continuum intensity, predicting magnetic flux rather than continuum intensity is more important for space weather forecasting. This lag between the decrease in continuum intensity and the increase in magnetic flux is also evident when comparing the predictions of the two studies, where in the best case of \cite{kasapis2025prediction}, continuum intensity was predicted 36 hours ahead, whereas for magnetic flux, the best prediction was 10 hours ahead for AR13183 (Table~\ref{table:best_vanillalstm_hybrid}). 

Through systematic comparison of two LSTM-based architectures, a stacked LSTM (\texttt{MagFluxLSTM}) and an encoder-decoder architecture (\texttt{MagFluxEnc-Dec}), we demonstrate that the simpler stacked LSTM with the hybrid loss formulation achieves superior predictive performance for this task. Extensive hyperparameter optimization revealed that 89\% of the top-100 model configurations utilized the stacked \texttt{MagFluxLSTM} architecture, indicating strong convergence to this simpler, more robust approach for magnetic flux prediction. This work presents the first ML model trained for predicting magnetic flux associated with the emergence of solar active regions. Our results show that the \texttt{MagFluxLSTM} model with a hybrid loss can predict the emergence of magnetic flux 3-10 hours before it is observed, in an operational-type setting. The increase of magnetic flux associated with the emergence of the respective ARs was successfully predicted, in an operational-type setting, in three out of five cases. 

The current model operates on individual tiles without explicit spatial-correlation modeling. Incorporating additional components, such as CNNs, into \texttt{MagFluxLSTM} to capture spatial emergence patterns may improve performance for extended AR complexes. While challenges remain in scaling to larger datasets and integrating more diverse inputs, our approach represents a step toward operational forecasting tools that combine physical insight with data-driven methods. Several avenues for future development are promising: integrating vector magnetic field observations could enhance prediction accuracy by providing a more complete characterization of magnetic field topology, and transfer learning approaches could enable application to related prediction problems. Beyond immediate forecasting applications, this work demonstrates the potential of deep learning approaches to extract predictive information from subtle pre-emergence signatures in solar observations. By balancing absolute flux prediction with derivative sensitivity through the hybrid loss we introduced, the \texttt{MagFluxLSTM} model captures both the magnitude and temporal evolution of flux emergence, enabling more accurate identification of the magnetic flux onset that can be associated with AR emergence. Furthermore, the model's success in capturing pre-emergence signatures suggests potential applications to other solar activity prediction problems, such as solar flare prediction, filament formation forecasting, and coronal hole evolution modeling and others.


%

%

\section*{Open Research Section}
The model pipeline was implemented in PyTorch 2.7.0 on NVIDIA GPUs. Ray Tune module and ReduceLROnPlateau, TrialPlateauStopper, ASHAScheduler, Hyperopt implementations were used throughout the code. The SolARED capabilities and ML-ready data are available for download from \url{https://sun.njit.edu/sarportal/}. The acoustic power maps analyzed in this work are available from the corresponding author upon request. The code used to generate and process the dataset, following the methodology described in the paper, is available at \url{https://github.com/jonastirona/SAR_EMERGENCE_RESEARCH/tree/flip_flux_and_intensity}.

%
\begin{acks}
This work was supported by the New Jersey Institute of Technology (NJIT) Grace Hopper AI Research Institute (GHAIRI) Seed Grant; the NASA AI/ML HECC Expansion Program, and the NASA grants 23-HGIO23\textunderscore2-0077, 20-HSR20\textunderscore2-0037, 80NSSC19K0630, 80NSSC19K0268, 80NSSC20K1870, and 80NSSC22M0162. We also thank the computing resources provided by the High Performance Computing (HPC) facility at NJIT.
\end{acks}

%
%
%
%
%
%
%

%
%
%
%
%
%

\bibliographystyle{spr-mp-sola}
\bibliography{agusample}

%

\appendix

\section{Hyperparameter Optimization}
\label{appendix:hyperopt}

Hyperparameter optimization was performed using Ray Tune using the Hyperopt search algorithm
and the Asynchronous Successive Halving Algorithm (ASHA) for early-stopping of underperforming trials.
The optimization objective was set to minimize the validation root-mean-square error (RMSE) of magnetic flux prediction. 
Maximum 100 epochs were allowed per trial, but trials could be terminated when their RMSE plateaued for 10 consecutive epochs as determined by the \texttt{TrialPlateauStopper} or by ASHA.
The complete search space and tuning configuration can be found in Listing~\ref{lst:hyperopt_code}. 

\begin{lstlisting}[language=Python, caption={Ray Tune + Hyperopt search configuration for LSTM hyperparameter optimization.}, label={lst:hyperopt_code}]
search_space = {
    "learning_rate": hp.loguniform("learning_rate", log(1e-5), log(1e-2)),
    "hidden_size": hp.choice("hidden_size", [2, 4, 8, 16, 32, 64, 128]),
    "num_layers": hp.choice("num_layers", [1, 2, 3, 4]),
    "dropout": hp.choice("dropout", [0, 0.1, 0.2, 0.3]),
    "batch_size": hp.choice("batch_size", [32, 64]),
    "weight_decay": hp.loguniform("weight_decay", log(1e-6), log(1e-3)),
    "n_epochs": 100,
    # Dataset
    "shuffle": hp.choice("shuffle", [True, False]),
    # Model architecture | Conditional Search Space
    "model": hp.choice(
        "model_branch",
        [
            {
                "model": "\texttt{MagFluxLSTM}",
            },
            {
                "model": "\texttt{MagFluxEnc-Dec}",
                "teacher_forcing_ratio": hp.choice(
                    "teacher_forcing_ratio", [0, 0.1, 0.15, 0.25, 0.5]
                ),
            },
        ],
    ),
    "lossFn": hp.choice(
        "lossFn_branch",
        [
            {
                "lossFn": "hybrid",
                "alpha": hp.choice("alpha", [0.1, 0.3, 0.5, 0.7, 0.9]),
            },
            {"lossFn": "value"},
        ],
    ),
}

# Scheduler to early-stop bad trials
scheduler = ASHAScheduler(
    metric="RMSE",
    mode="min",
    grace_period=15,  # Min epochs before a trial can be stopped
    reduction_factor=2,
)

# Search algorithm
search_alg = HyperOptSearch(space=search_space, metric="RMSE", mode="min")

early_stopper = TrialPlateauStopper(
    metric="RMSE",
    mode="min",
    grace_period=10,  # Number of epochs to wait for improvement
)

# Set up the Tuner
ray.init(num_cpus=32, num_gpus=1, include_dashboard=False, _temp_dir="/tmp/ray")
train_ref = ray.put(tensor_train)
val_ref = ray.put(tensor_val)
tuner = tune.Tuner(
    tune.with_resources(
        tune.with_parameters(
            main,
            train_ref=train_ref,
            val_ref=val_ref,
        ),
        {"gpu": 1 / 32, "cpu": 1},
    ),
    tune_config=tune.TuneConfig(
        num_samples=parse_args()[
            "sample_size"
        ],  # Number of different hyperparameter combinations to try
        scheduler=scheduler,
        search_alg=search_alg,
    ),
    run_config=ray.train.RunConfig(
        name="lstm_hyperparameter_search",
        stop=early_stopper,  # Max epochs per trial
    ),
)
\end{lstlisting}

\section{Additional Results} 
\label{appendix:results_denormalized}
\begin{figure}[H]
    \centering
    \includegraphics[width=\textwidth, height=\textheight, keepaspectratio]{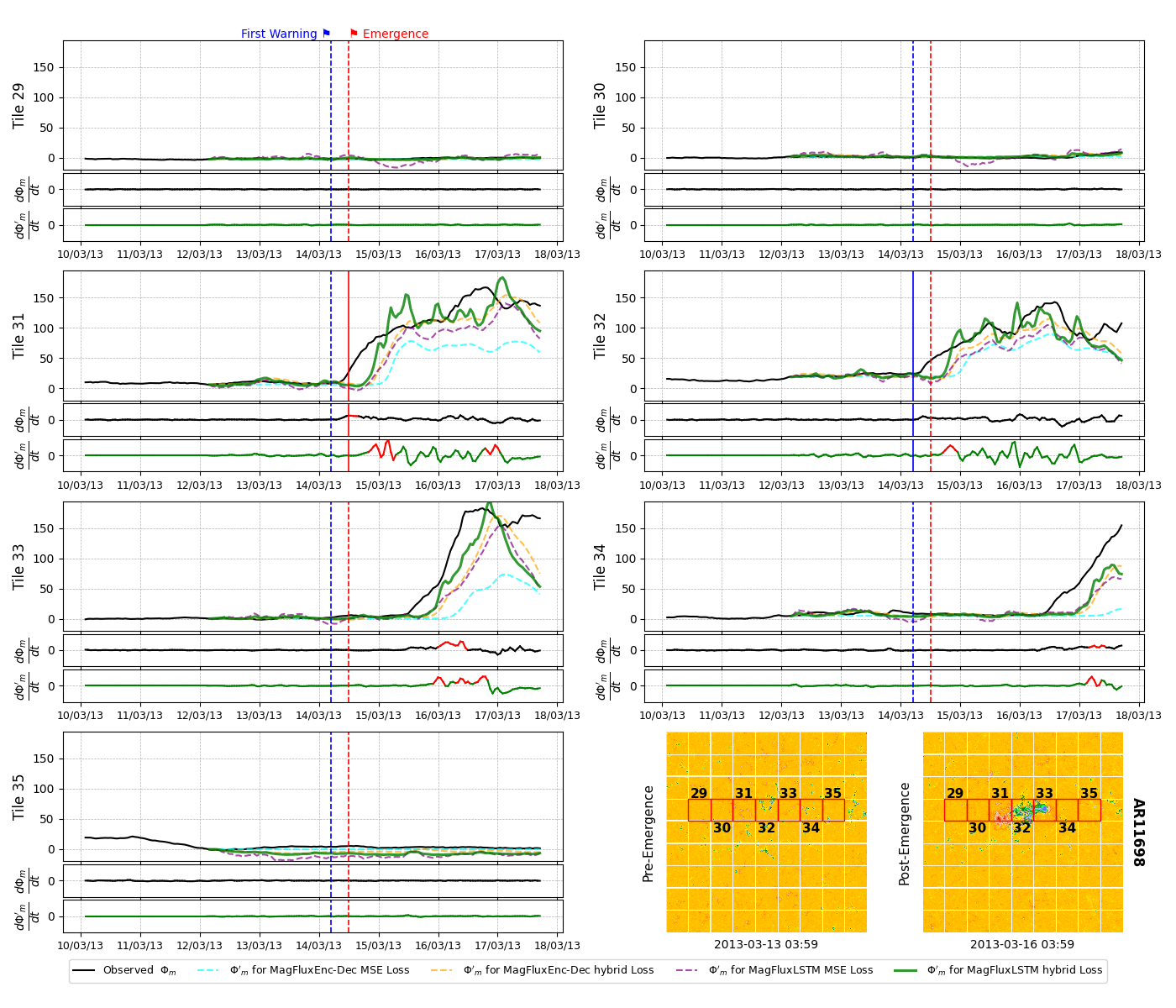}
    \caption{Same as Figure~\ref{fig:results} for AR11698.}
    \label{fig:results11698}
\end{figure}


\begin{figure}[H]
    \centering
    \includegraphics[width=\textwidth, height=\textheight, keepaspectratio]{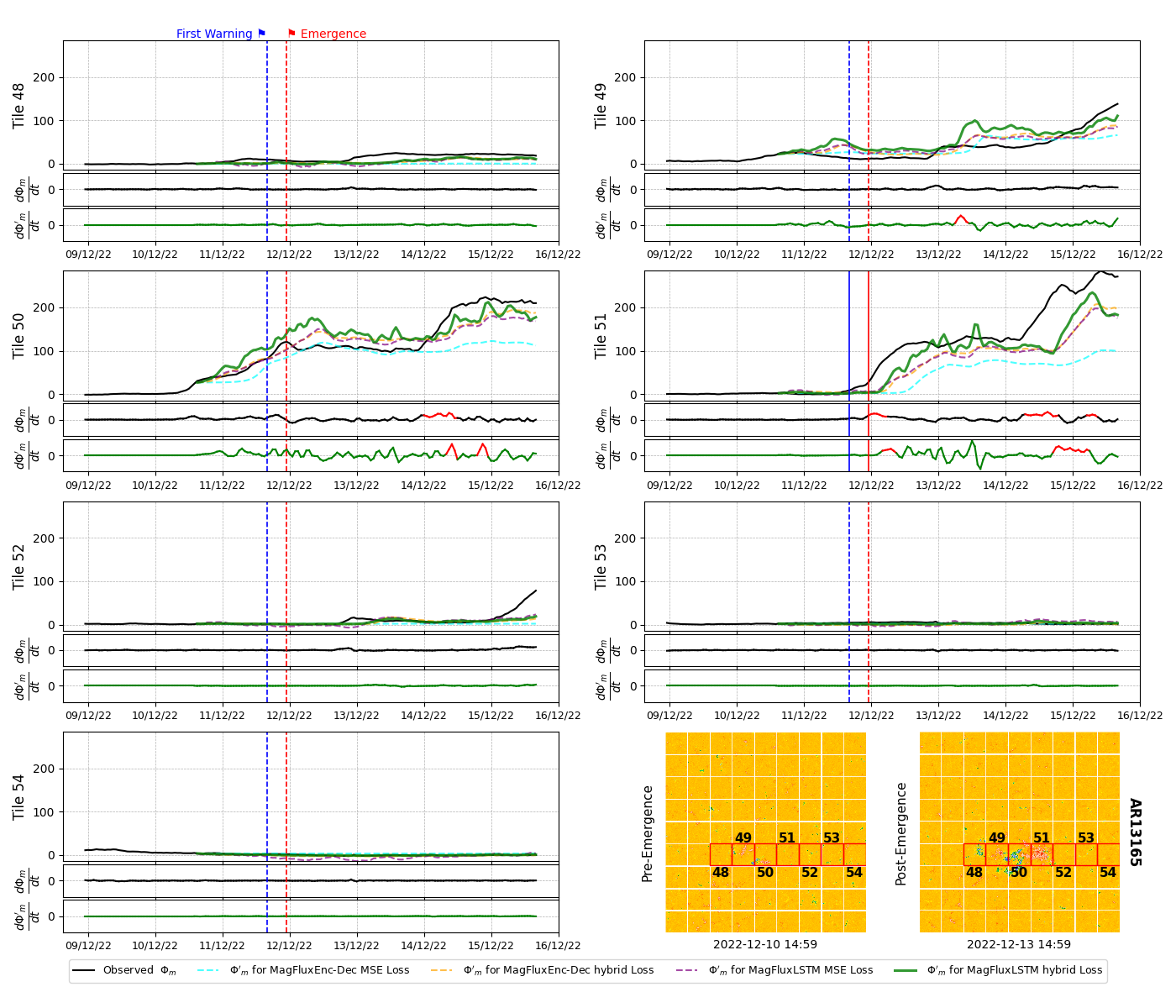}
    \caption{Same as Figure~\ref{fig:results} for AR13165.}
    \label{fig:results13165}
\end{figure}


\begin{figure}[H]
    \centering
    \includegraphics[width=\textwidth, height=\textheight, keepaspectratio]{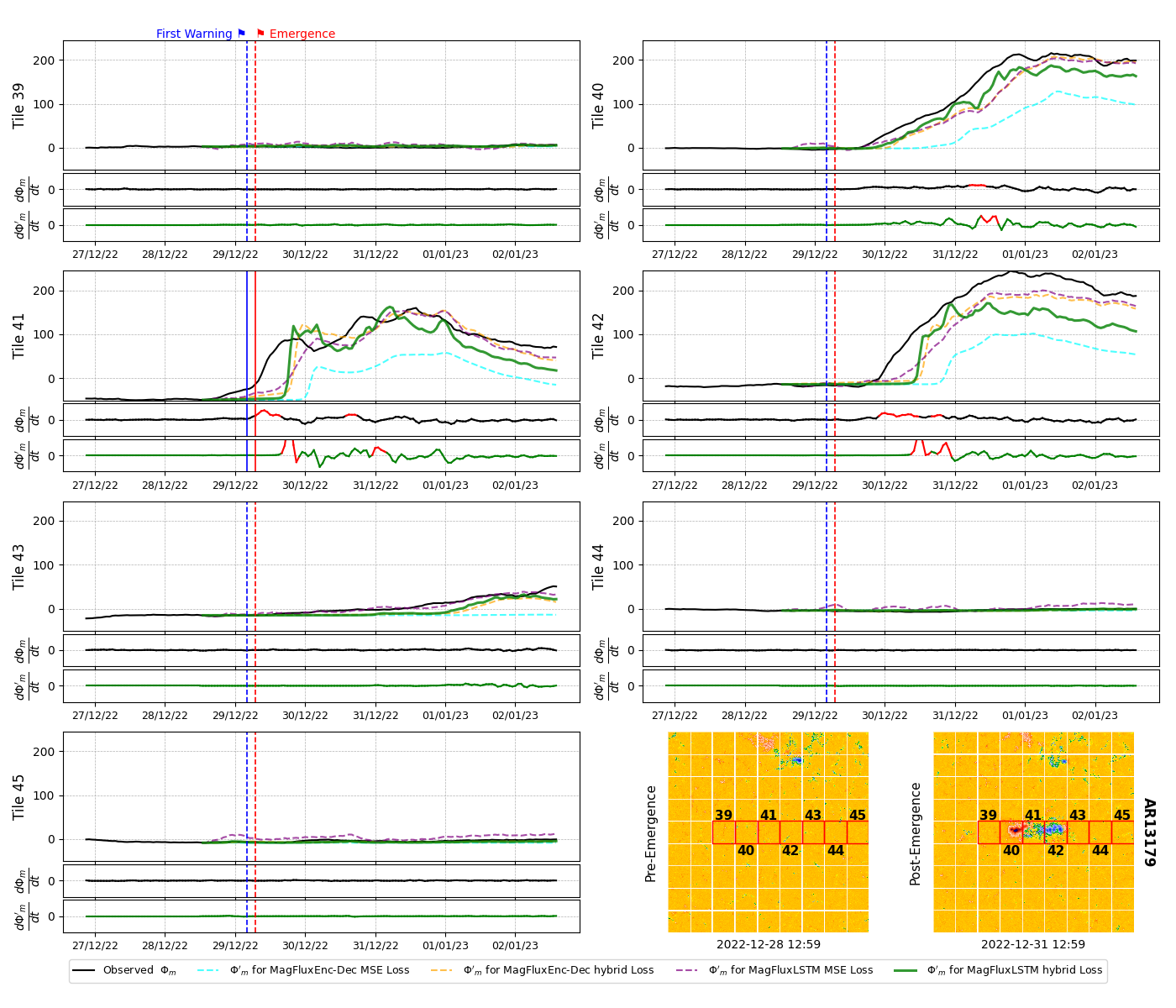}
    \caption{Same as Figure~\ref{fig:results} for AR13179.}
    \label{fig:results13179}
\end{figure}

\vspace{1em} 

\begin{figure}[H]
    \centering
    \includegraphics[width=\textwidth, height=\textheight, keepaspectratio]{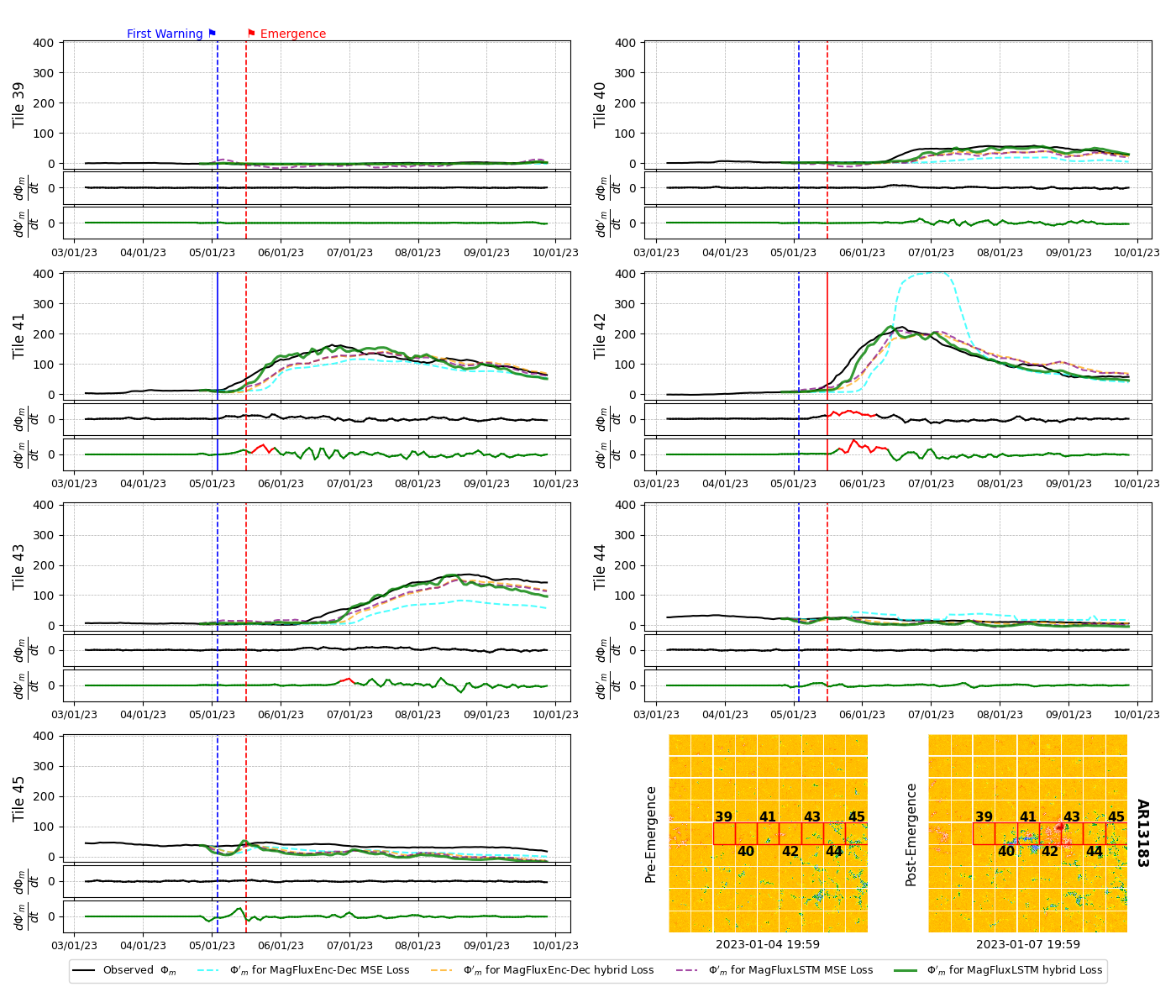}
    \caption{Same as Figure~\ref{fig:results} for AR13183.}
    \label{fig:results13183}
\end{figure}

\end{document}